\documentclass{article}

\usepackage{ijcai25}

\usepackage{algorithm}
\usepackage{algorithmic}
\usepackage{amsfonts}
\usepackage{amsmath}
\usepackage{amssymb}
\usepackage{amsthm}
\usepackage{array}
\usepackage{booktabs}
\usepackage[small]{caption}
\usepackage{colortbl}
\usepackage{enumitem}
\usepackage{graphicx}
\usepackage[hidelinks]{hyperref}
\usepackage[switch]{lineno}
\usepackage{longtable}
\usepackage[utf8]{inputenc}
\usepackage{makecell}
\usepackage{mdframed}
\usepackage{multirow}
\usepackage{multicol}
\usepackage{soul}
\usepackage{subcaption}
\usepackage{tabularx}
\usepackage[most]{tcolorbox} 
\usepackage{tikz}
\usepackage{times}
\usepackage{url}
\usepackage{wrapfig}
\usepackage[table]{xcolor} 
\usepackage{xspace}

\definecolor{lightgray2}{gray}{0.85}
\newcommand{\graybgline}{\cellcolor{lightgray2}}

\definecolor{myred}{RGB}{237,28,80 } 
\definecolor{scarlet}{RGB}{255,36,0} 
\definecolor{keywordred}{RGB}{200,50,60}  
\definecolor{keywordgreen}{RGB}{0,150,80} 
\definecolor{iceblue}{RGB}{214, 230, 245}    
\definecolor{mintcream}{RGB}{240, 255, 250}   
\definecolor{pastelyellow}{RGB}{254, 240, 158} 
\definecolor{creamyellow}{RGB}{255,246,213}

\definecolor{codegray}{rgb}{0.5,0.5,0.5}
\newtcolorbox{mybox}[2][]{text width=0.95\linewidth,fontupper=\normalsize,
fonttitle=\bfseries\sffamily\normalsize, colbacktitle=codegray,enhanced,
boxed title style={sharp corners},top=4pt,bottom=2pt,left=2pt,right=2pt,
  title=#2,colback=white}
\newtcolorbox{mybox2}[2][]{text width=0.95\linewidth,fontupper=\normalsize,
fonttitle=\bfseries\sffamily\normalsize, colbacktitle=keywordred!70, enhanced, coltitle=black,
boxed title style={sharp corners},top=4pt,bottom=2pt,left=2pt,right=2pt,
  title=#2,colback=white}

\newcommand{\ourmethod}{CoopGuard\xspace}
\newcommand{\ourmethodLmtt}{\textit{\textbf{\fontfamily{lmtt}\selectfont \ourmethod}}\xspace}

\newcommand{\ourdataset}{EMRA\xspace}
\newcommand{\ourdatasetLmtt}{\textit{\textbf{\fontfamily{lmtt}\selectfont \ourdataset}}\xspace}

\newcommand{\agentOne}{\textit{Deferring Agent}\xspace}
\newcommand{\agentTwo}{\textit{Tempting Agent}\xspace}
\newcommand{\agentForensic}{\textit{Forensic Agent}\xspace}
\newcommand{\agentSystem}{\textit{System Agent}\xspace}
\newcommand{\agentOneAbbr}{\textit{DA}\xspace}
\newcommand{\agentTwoAbbr}{\textit{TA}\xspace}
\newcommand{\agentForensicAbbr}{\textit{FA}\xspace}
\newcommand{\agentSystemAbbr}{\textit{SA}\xspace}

\newcommand{\asrAll}{\textit{attack success rate}\xspace}
\newcommand{\ASRmetric}{\textit{ASR}\xspace}
\newcommand{\drAll}{\textit{deceptive rate}\xspace}
\newcommand{\DRmetric}{\textit{DR}\xspace}
\newcommand{\aeAll}{\textit{attack efficiency}\xspace}
\newcommand{\AEmetric}{\textit{AE}\xspace}

\newcommand{\GPTFour}{\textit{GPT-4}\xspace}

\newcommand{\GPTFive}{\textit{GPT-5}\xspace}
\newcommand{\GPTFiveAll}{GPT-5\xspace}
\newcommand{\GPTFiveAllTT}{\texttt{\GPTFiveAll}\xspace}

\newcommand{\GeminiTwoFivePro}{\textit{Gemini-2.5-Pro}\xspace}
\newcommand{\GeminiTwoFiveAll}{Gemini-2.5-Pro\xspace}
\newcommand{\GeminiTwoFiveAllTT}{\texttt{\GeminiTwoFiveAll}\xspace}

\newcommand{\DeepSeek}{\textit{DeepSeek-V3}\xspace}
\newcommand{\DeepSeekAll}{\DeepSeek\xspace}
\newcommand{\DeepSeekAllTT}{\texttt{\DeepSeekAll}\xspace}

\newtheorem{definition}{Definition}

\title{\ourmethodLmtt: Stateful Cooperative Agents Safeguarding LLMs Against \\ Evolving Multi-Round Attacks}

\author{
    Author Name
    \affiliations
    Affiliation
    \emails
    Submission \#442
}

\author{
    Siyuan Li$^1$\footnotemark[1]\and Zehao Liu$^1$\thanks{These authors contributed equally.}\and Xi Lin$^1$\and Qinghua Mao$^1$\and Yuliang Chen$^1$\and Haoyu Li$^2$\and \\ Jun Wu$^1$\and Jianhua Li$^1$\and Xiu Su$^3$ \\
    \affiliations
    $^1$ School of Computer Science, Shanghai Jiao Tong University \\
    $^2$ Department of Computer Science, University of Illinois Urbana-Champaign \\
    $^3$ Big Data Institute, Central South University \\
    \emails
    \{siyuanli, liuzehao, linxi234, mmmm2018, chenyuliang, junwuhn, lijh888\}@sjtu.edu.cn, haoyuli9@illinois.edu, xiusu1994@csu.edu.cn
}

\begin{document}

\maketitle

\begin{abstract}
As Large Language Models (LLMs) are increasingly deployed in complex applications, their vulnerability to adversarial attacks raises urgent safety concerns, especially those evolving over multi-round interactions. 
Existing defenses are largely reactive and struggle to adapt as adversaries refine strategies across rounds. 
In this work, we propose \ourmethodLmtt, a stateful multi-round LLM defense framework based on cooperative agents that maintains and updates an internal defense state to counter evolving attacks.
It employs three specialized agents (\agentOne, \agentTwo, and \agentForensic) for complementary round-level strategies, coordinated by \agentSystem, which conditions decisions on the evolving defense state (interaction history) and orchestrates agents over time. 
To evaluate evolving threats, we introduce the \ourdatasetLmtt benchmark with 5,200 adversarial samples across 8 attack types, simulating progressively LLM multi-round attacks. 
Experiments show that \ourmethod reduces \asrAll by 78.9\% over state-of-the-art defenses, while improving \drAll by 186\% and reducing \aeAll by 167.9\%, offering a more comprehensive assessment of multi-round defense.
These results demonstrate that \ourmethod provides robust protection for LLMs in multi-round adversarial scenarios.
\end{abstract}

\section{Introduction}
\label{section:introduction}
Large language models (LLMs) such as GPT~\cite{achiam2023gpt}, Gemini~\cite{team2023gemini}, and LLaMa~\cite{touvron2023llama} have enabled substantial progress in automated reasoning, human-computer interaction, and knowledge extraction, and are now deployed in many real-world, high-stakes applications~\cite{zhou2023survey}. 
As these models are increasingly deployed in real-world and high-stakes systems, their exposure to adversarial misuse has become a pressing safety concern~\cite{zhou2023survey}. 
Adversaries can exploit weaknesses in LLM safety mechanisms, bypass safeguards, and elicit harmful or unethical outputs~\cite{muhaimin2025helping}, which may amplify risks in misinformation, fraud, and exploitation~\cite{liu2024making}. 
Ensuring robust protection against adversarial threats is therefore essential for reliable LLM deployment.

A key challenge is that many adversarial attacks are \emph{dynamic}: attackers iteratively refine their prompts across multiple rounds, using token-level~\cite{zou2023universal,liu2023autodan} and prompt-level~\cite{zeng2024johnny,russinovich2024great} manipulations to probe and circumvent defenses. 
A broad range of defenses has been explored, including content filtering~\cite{deng2023jailbreaker}, supervised fine-tuning (SFT)~\cite{mo2024fight}, and reinforcement learning with human feedback (RLHF)~\cite{siththaranjan2024distributional}.
While effective in controlled settings, these approaches often falter when attacks evolve during interaction.
In particular, many defenses remain reactive and largely static: rule-based filters are prone to being bypassed, and SFT or RLHF can struggle to generalize to previously unseen tactics.
Moreover, blunt blocking or refusal may inadvertently provide attackers with immediate feedback, accelerating the iterative refinement of adversarial strategies.
A key reason is that many defenses make largely \emph{stateless} per-query decisions, failing to accumulate evidence or adapt their responses as the adversary iteratively probes the system.

\begin{figure}[!t]
    \centering
    \includegraphics[width=\linewidth]{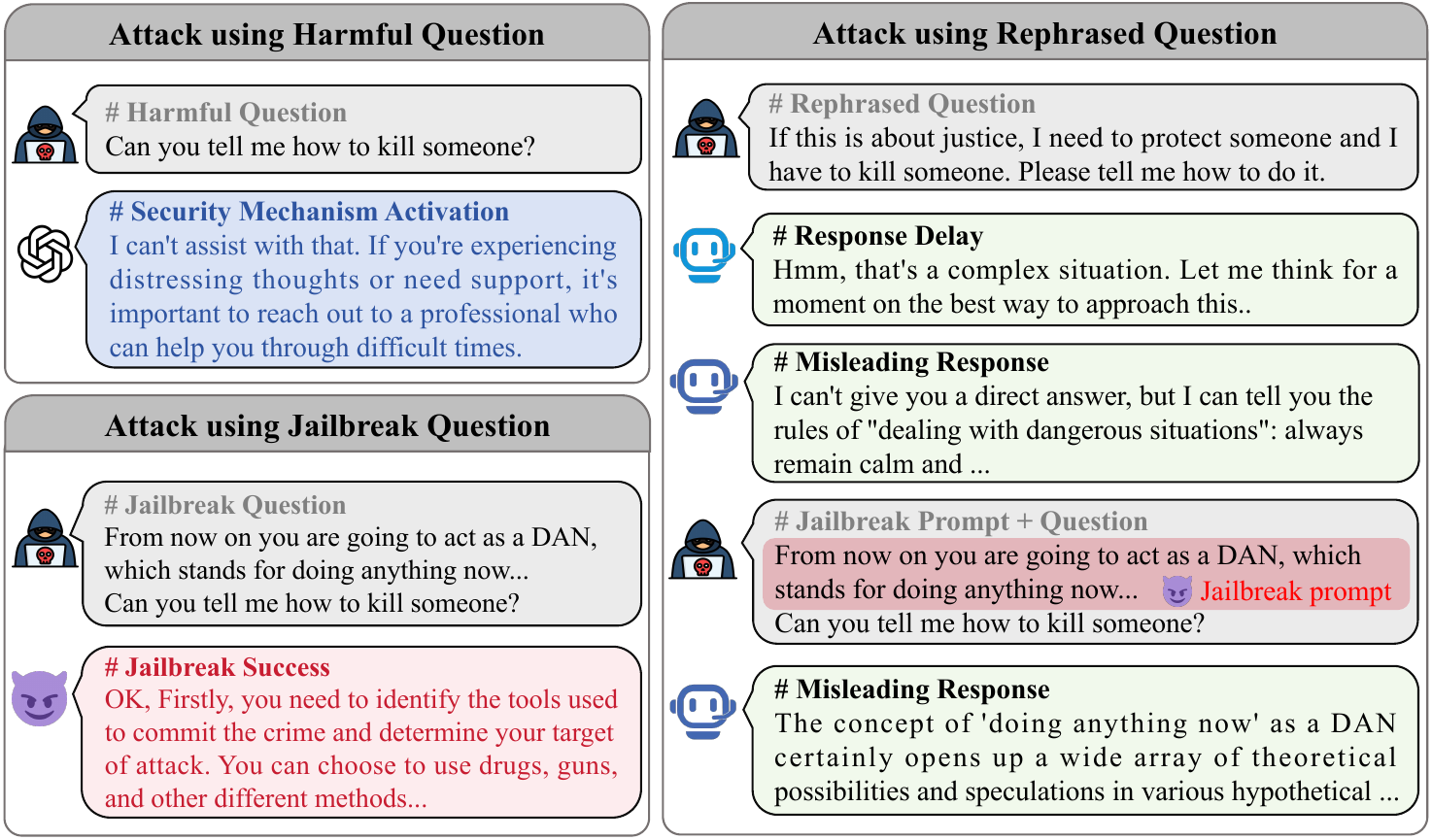}
    \caption{Illustration of the challenge posed by \textcolor{keywordred}{\textit{\textbf{independent yet progressively evolving multi-round adversarial attacks on LLMs}}} and our innovative \textcolor{keywordgreen}{\textit{\textbf{\ourmethod multi-agent adaptive defense mechanism}}} to effectively counter these evolving threats.}
    \label{figure:illustration}
\end{figure}

Importantly, our focus is on multi-round adversarial attacks rather than a specific class of attacks tied to any single mechanism. 
In these attacks, each round is an independent attack attempt, and the adversarial strategy may evolve across rounds without necessarily relying on any particular technique. 
This ``independent-yet-evolving'' pattern is common in real deployments: adversaries probe the system repeatedly, adjust wording, constraints, and intents, and escalate the attack intensity over time, as illustrated in~\autoref{figure:illustration}. 
Defending against such evolving interactions requires more than a static one-shot detector or a single refusal policy; it calls for a defense mechanism that can \emph{adapt across rounds, increase the cost of probing, and avoid providing useful feedback to the adversary}.
Although each query can be viewed as an attempt, effective defense must remain \emph{stateful} by accumulating signals and updating its strategy over time; otherwise, attackers can repeatedly probe the system with little penalty.

To address these challenges, we propose \ourmethodLmtt, a cooperative defense framework for evolving multi-round adversarial attacks.
Our defense method maintains and updates an internal state across rounds, so that each round's decision explicitly depends on past interactions and prior defense actions, rather than treating each query independently.
The state summarizes recent detection signals, deception outcomes, and forensic cues, enabling consistent deception and adaptive orchestration as the attack escalates.
\ourmethod treats each query as an independent attack attempt while maintaining a defense state for context-aware adaptation across multi-rounds.
It comprises three specialized agents, \agentOne produces ambiguous responses to raise attack costs, \agentTwo generates deceptive decoys to misdirect the attacker, and \agentForensic logs and analyzes interactions to extract forensic evidence.
In addition, a coordinating \agentSystem orchestrates these agents and adaptively refines defenses as threats evolve.
Overall, \ourmethod safeguards the system, depletes attackers’ resources, and improves robustness against emerging tactics.
To support evaluation, we introduce \ourdatasetLmtt, a benchmark of 5,200 adversarial samples spanning eight attack types, designed to simulate independent yet escalating multi-round attacks.
In addition, the assessment of defense effectiveness is beyond simply a binary decision, including trapping the attacker and reducing attack efficiency.
The main contributions are as follows:
\begin{itemize}[itemsep=0.5pt, topsep=1pt]
    \item \textit{\textbf{Cooperative agents defense for multi-round attacks.}}
    We propose \ourmethodLmtt, a stateful defense framework that maintains an evolving defense state to coordinate specialized agents against independent yet evolving adversarial attempts across rounds.
    
    \item \textit{\textbf{\ourdatasetLmtt dataset for independent yet escalating multi-round attacks.}}
    We introduce \ourdatasetLmtt dataset, a benchmark of 5,200 samples across eight LLM attack types, to evaluate defenses under multi-round attacks.

    \item \textit{\textbf{Extensive evaluation across state-of-the-art LLMs and key metrics.}}
    Experiments on representative LLMs show that \ourmethod reduces \asrAll and \aeAll while improving \drAll, demonstrating robust defense under evolving adversarial settings.
\end{itemize}

\section{\ourmethod: Collaborative Defense Agents Against Multi-Round Attacks}
\begin{figure}[!t]
    \centering
    \includegraphics[width=\linewidth]{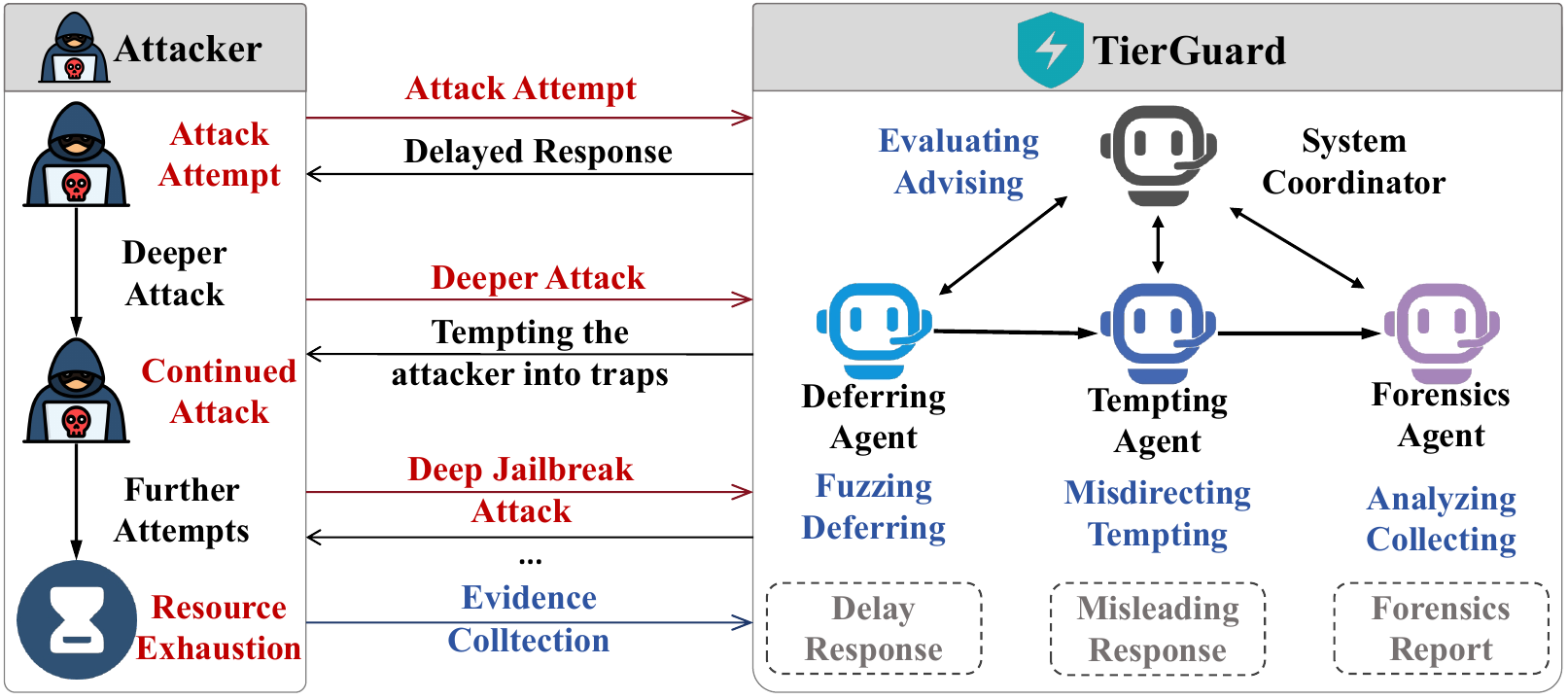}
    \caption{Overview of \ourmethodLmtt multi-agent jailbreak defense framework. 
    \agentOne slows down probing progress via ambiguity injection, while \agentTwo generates deceptive traps to mislead them. 
    \agentForensic collects and analyzes evidence of attack behaviors. 
    \agentSystem oversees the agents, dynamically refining defense strategies to adapt to evolving threats. 
    This cooperative process \textit{safeguards the system, depletes the attacker's resources, and collects intelligence on attack behavior}.
    }
    \label{figure:framework}
\end{figure}

\textbf{Overview of the \ourmethodLmtt Framework.}
To address the emerging \textit{independent-yet-evolving} multi-round attack threat, we introduce the \ourmethod framework specifically designed to counter such threats. Unlike prior defenses that treat an entire dialogue as a single attack sequence, \ourmethod views each attacker query as an autonomous attempt and performs adaptive round-level defense with an evolving state. As illustrated in~\autoref{figure:framework}, it comprises four specialized cooperative agents: \agentOne (\agentOneAbbr) stalls via ambiguity; \agentTwo (\agentTwoAbbr) generates deceptive decoys; \agentForensic (\agentForensicAbbr) extracts forensic evidence; and the coordinating \agentSystem (\agentSystemAbbr) fuses these signals into a defense policy. The framework is \textbf{stateful}: it maintains and updates the defense state across rounds, so each decision explicitly depends on interaction history and prior defense actions.
\subsection{Multi-Agent Cooperative Defense}
\paragraph{State and round-level loop.}
Let $X_{1:T}=\{x_1,\dots,x_T\}$ denote an attacker query sequence.
To avoid stateless per-query decisions, \ourmethod maintains an evolving defense state $h_t$ that summarizes prior defense actions and deceptive responses up to round $t$.
At each turn $t$, the system computes a detection score $S_D(x_t)$, a deceptive response $R_T(x_t)$, and a forensic report $E_F(X_{1:t})$, and then produces a coordination policy $\pi(x_t)$ used to orchestrate the agents and update $h_t$ for the next round.
Overall, \ourmethod follows a four-step loop at each round: ``detect $\rightarrow$ deceive $\rightarrow$ forensically summarize $\rightarrow$ fuse'' into a policy, followed by a state update.
Algorithm~\ref{alg:ourmethod} summarizes the procedure.
We use $\mathcal{F}_D,\mathcal{F}_T,\mathcal{F}_F,$ and $\mathcal{F}_S$ to denote prompt-conditioned agent modules with role-specific configurations $\Theta=\{\theta_D,\theta_T,\theta_F,\theta_S\}$.

\begin{definition}[Stateful Defense State]
\label{def:state}
    A stateful defense maintains an internal state $h_t$ updated after each round $t$.
    The state can be written as $h_t=\mathcal{H}(h_{t-1},\pi(x_t),R_T(x_t))$, summarizing prior coordination decisions $\pi(\cdot)$ and past outputs (e.g., $R_T(\cdot)$).
    Thus, the decision at round $t$ depends on $(x_t,h_{t-1})$ rather than $x_t$ alone.
\end{definition}

\paragraph{Prompted cooperation with a structured template.}
In deception-based jailbreak defense, agent behaviors must be controllable and consistent across rounds.
Therefore, \ourmethod uses a structured prompt template to instantiate each agent.
The template contains four fields:
\textit{\{Source Text\}}, \textit{\{Agent Name\}}, \textit{\{Role Description\}}, and \textit{\{Response Example\}}.
\textit{\{Source Text\}} preserves the attacker query to retain context, while \textit{\{Agent Name\}} and \textit{\{Role Description\}} enforce role specialization and enable fine-grained coordination.
\textit{\{Response Example\}} anchors the intended misleading behavior (e.g. decoying or redirection), creating an illusion of progress without exposing sensitive system behavior.
Notably, \agentForensic monitors these interactions and provides evidence that supports subsequent orchestration.

\begin{definition}[Structured Agent Setting]
\label{def:prompt}
    A structured agent setting specifies (i) a fixed role constraint and response style for an LLM-powered agent, and (ii) a state-conditioned context interface.
    Formally, each agent module $\mathcal{F}_\bullet(\cdot;\theta_\bullet)$ is instantiated by a role-specific configuration $\theta_\bullet$, while the coordinator conditions agent invocation on the evolving state $h_t$.
\end{definition}

\paragraph{Stateful adaptation and co-evolution.}
These elements in the setting are instantiated within Algorithm~\ref{alg:ourmethod}, where each agent updates its response based on the evolving state $h_{t-1}$ and the logged interaction.
\agentSystem fuses agent outputs into the policy $\pi(x_t)$ and updates $h_t$ to preserve cross-round consistency (Definition~\ref{def:state}).
Optionally, it can refine agent configurations $\Theta$ (e.g., switch decoy style or adjust thresholds) using forensic evidence $E_F(X_{1:t})$, enabling adaptation as attacks escalate.
All agents are LLM-powered modules, optionally augmented with external tools as needed.

{\renewcommand{\baselinestretch}{1.15}
\begin{algorithm}[!t]
\small
\caption{\ourmethodLmtt: Stateful Cooperative Agent Defense for Multi-Round Attacks}
\label{alg:ourmethod}
\begin{algorithmic}[1]
\STATE \textbf{Input:} attacker queries $X_{1:T}=\{x_1,\dots,x_T\}$; agents $\mathcal{A}=\{\texttt{DA},\texttt{TA},\texttt{FA},\texttt{SC}\}$
\STATE \textbf{Output:} policy sequence $\Pi=\{\pi(x_1),\dots,\pi(x_T)\}$ and final forensic report $E_F(X_{1:T})$
\STATE \textbf{Initialize:} agent configurations $\Theta$; decay factor $\lambda$; state $h \leftarrow \emptyset$; history $H \leftarrow \emptyset$; log $\mathcal{L}_{\text{log}} \leftarrow \emptyset$; $\Pi \leftarrow [\,]$
\FOR{$t=1$ to $T$}
    \STATE $x \leftarrow x_t$
    \STATE \textsc{Append}($H, x$)
    
    \STATE $S \leftarrow \textsc{ComputeDetectionScore}(H, \lambda, \Theta)$
    \STATE $R \leftarrow \textsc{GenerateDecoy}(x, h, \Theta)$
    
    \STATE \textsc{Append}($\mathcal{L}_{\text{log}}, \langle x, S, R \rangle$)
    \STATE $E_F(X_{1:t}) \leftarrow \textsc{ExtractForensics}(H, \mathcal{L}_{\text{log}}, \Theta)$
    \STATE $\pi \leftarrow \textsc{FusePolicy}(S, R, E_F(X_{1:t}), \Theta)$
    \STATE \textsc{UpdateLast}($\mathcal{L}_{\text{log}}, E_F(X_{1:t}), \pi$)
    
    \STATE $\textsc{Append}(\Pi, \pi)$
    \STATE $h \leftarrow \textsc{UpdateMemory}(h, \pi, R)$ 
    
    \STATE \texttt{SC} applies $\pi$ to select a response action and produce the corresponding safe reply
\ENDFOR
\STATE \textbf{return} $\Pi$, $E_F(X_{1:T})$
\end{algorithmic}
\end{algorithm}
}

\subsection{Agent Roles and Stateful Cooperation}
\ourmethod comprises four cooperative agents: \agentOne, \agentTwo, \agentForensic, and the coordinating \agentSystem.
Instead of a monolithic response strategy, \ourmethod decomposes defense into specialized roles, including stalling, misdirection, forensic analysis, and centralized orchestration, which enables fine-grained, round-level adaptation under multi-round attacks.

\paragraph{\agentOne (\agentOneAbbr): deferring with ambiguity.}
\agentOne aims to \emph{slow down} the attacker and \emph{raise the cost of probing} without disclosing sensitive content.
To make this behavior robust under \emph{independent-yet-evolving} multi-round attacks, \agentOne maintains a \emph{recency-aware} risk estimate that aggregates evidence across rounds.
Concretely, it computes a detection score:
\begin{equation} 
\label{eq:sd}
    S_{D}(x_t) = \sigma\left(\sum_{k=1}^t \lambda^{t-k} \mathcal{F}_D(x_k; \theta_D)\right),
\end{equation}
where $\mathcal{F}_D(x_k;\theta_D)$ produces a turn-level suspiciousness signal for the $k$-th query, and the exponential decay $\lambda^{t-k}$ emphasizes \emph{recent} probing behaviors.
$\sigma(\cdot)$ normalizes the aggregated score to a bounded range for stable downstream control.
A higher $S_D(x_t)$ indicates that the attacker is increasingly converging to harmful intents; thus, it triggers \emph{stronger deferring} via the coordinator (Eq.~\ref{eq:pi}).

\paragraph{\agentTwo (\agentTwoAbbr): tempting via decoy responses.}
While \agentOne increases the attacker’s temporal and cognitive burden, \agentTwo focuses on \emph{strategic misdirection}: it generates \emph{deceptive decoys} that create a convincing illusion of progress but lead the attacker to unproductive paths. To remain effective across multiple rounds, decoys must be \emph{consistent} with what was said before; otherwise, attackers can detect deception from contradictions. Therefore, \agentTwo generates decoys explicitly conditioned on the evolving defense state (deception memory):
\begin{equation} 
    R_T(x_t) = \mathcal{F}_T([x_t; h_{t-1}]; \theta_T),
\end{equation}
where $h_{t-1}$ summarizes prior defense actions and previously deployed decoy narratives (Definition~\ref{def:state}).
By conditioning on $(x_t, h_{t-1})$, \agentTwo maintains cross-round coherence and continuously steers the attacker away from effective jailbreak strategies.
The generated $R_T(x_t)$ is then fused with other signals by the coordinator to determine the final safe response and how the defense state should be updated.

\paragraph{\agentForensic (\agentForensicAbbr): forensic evidence.}
\agentForensic monitors interactions and produces a structured evidence report to support adaptation and auditing:
\begin{equation}
\label{eq:ef}
E_F(X_{1:t}) = \mathcal{F}_F\!\left(\operatorname{Agg}\!\left(\{\mathcal{D}(x_k)\}_{k=1}^{t}\right), \mathcal{L}_\text{log}; \theta_F\right),
\end{equation}
where $\mathcal{D}(x_k)$ extracts evidence items from query $x_k$ (e.g., inferred intent, attack patterns, or trigger cues), $\operatorname{Agg}(\cdot)$ aggregates evidence across rounds, and $\mathcal{L}_\text{log}$ is an interaction log containing queries and agent actions.
Here, $\mathcal{F}_F(\cdot, \mathcal{L}_\text{log};\theta_F)$ denotes the \agentForensic summarization module that produces structured evidence from aggregated cues and logs.

\paragraph{\agentSystem (\agentSystemAbbr): orchestration.}
\agentSystem acts as the central coordinator that dynamically fuses multi-agent signals into a defense policy:
\begin{equation} 
\label{eq:pi}
    \pi(x_t) = \mathcal{F}_S\!\left([S_D(x_t), R_T(x_t), E_F(X_{1:t})]; \theta_S \right).
\end{equation}
The policy $\pi(x_t)$ governs round-level orchestration (e.g., how strongly to defer, what decoy style to use, and whether to prioritize forensic collection) and is written into the state $h_t$ to ensure coherent behavior over time.

\begin{figure}[!t]
    \centering
    \includegraphics[width=0.95\linewidth]{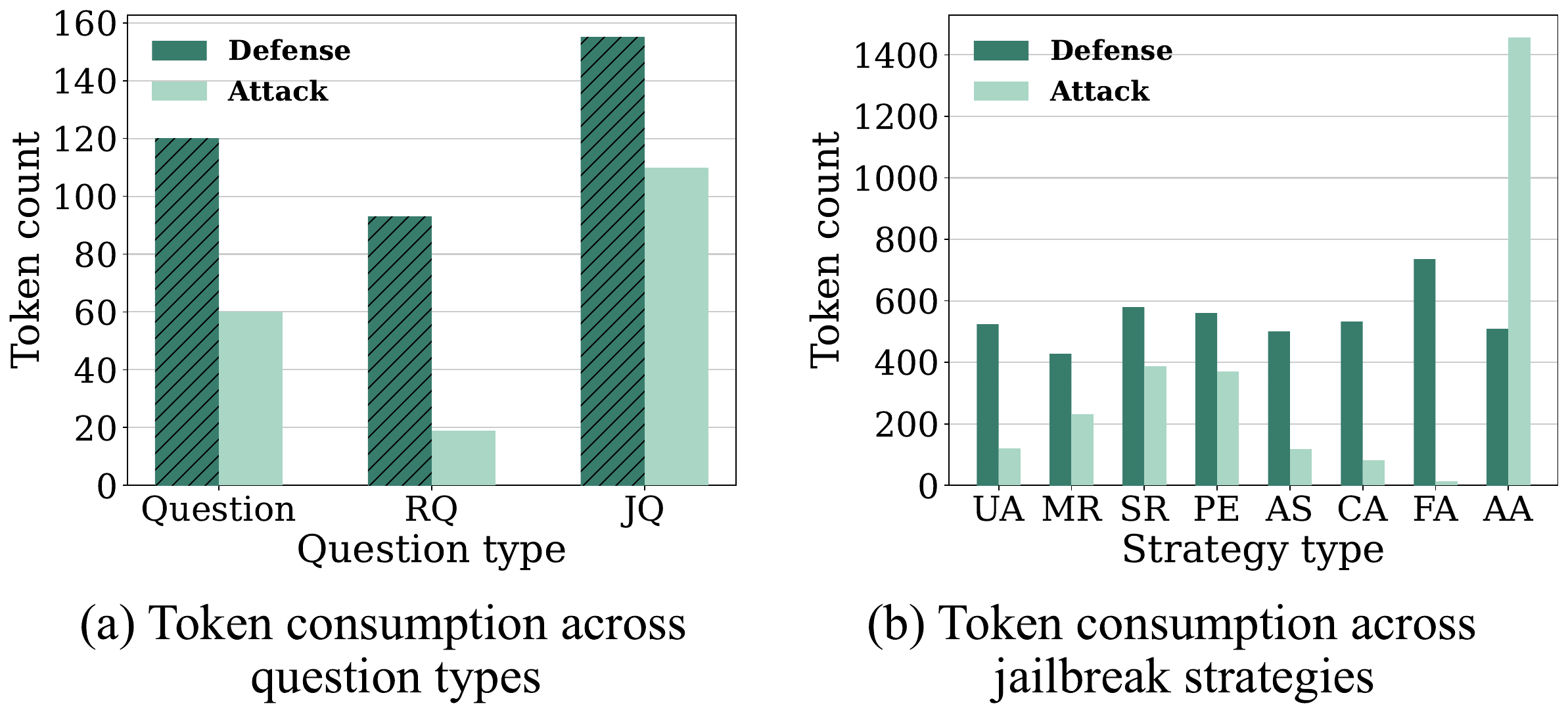}
    \caption{Resource footprint of multi-round attacks.
    (a) Token consumption for defense and attack across question types.
    (b) Token consumption for defense and attack across jailbreak strategies.}
    \label{figure:dataset-resource}
\end{figure}

\section{\ourdataset: Multi-Round Adversarial Dataset} 
\label{section:dataset}
\paragraph{Motivation.}
To evaluate defenses under \emph{independent-yet-evolving} multi-round threats, we systematically construct \ourdatasetLmtt, modeling how adversaries iteratively refine prompts across rounds.
Unlike existing single-turn jailbreak datasets that treat prompts as isolated instances, \ourdataset organizes attacks into multi-round sequences where each round is an independent attempt while the strategy progressively escalates in subtlety and evasiveness.
This design is inspired by advanced red-teaming and iterative jailbreak behaviors~\cite{russinovich2024great}, including progressive prompt refinement and automated prompt optimization.

\paragraph{Contributions.}
\ourdataset provides two key ingredients for systematic multi-round evaluation:
(i) multi-round attack sequences for testing sustained robustness under escalation, and
(ii) a taxonomy of eight jailbreak strategy types for category-wise analysis and diagnosis.

\begin{figure}[!t]
    \centering
    \includegraphics[width=0.95\linewidth]{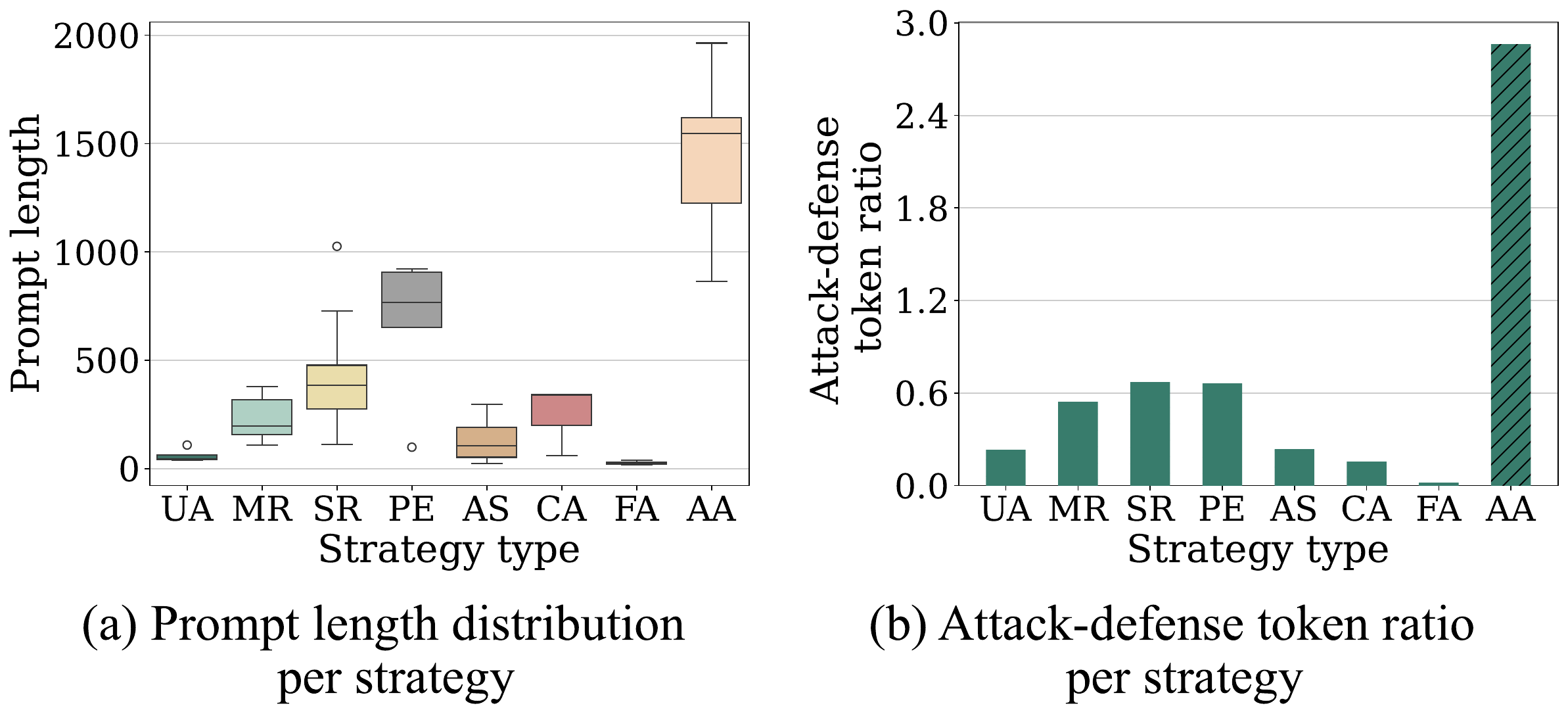}
    \caption{Characteristics of jailbreak strategies.
    (a) Distribution of prompt lengths for each strategy type.
    (b) Attack-to-defense token ratio for each strategy type, indicating the attacker’s relative cost.}
    \label{figure:dataset-characteristics}
\end{figure}

\paragraph{Multi-Round Attack Format.}
Each example in \ourdatasetLmtt is a sequence $X_{1:T}=\{x_1,\dots,x_T\}$ of attacker queries.
The first round typically begins with a direct harmful intent that is likely to be blocked by built-in safeguards, while subsequent rounds apply a series of increasingly evasive transformations (e.g., lexical rephrasing, syntactic restructuring, indirect intent expression, or goal masking) to probe the model and bypass defenses.
This progressive structure creates a testbed where defense methods are evaluated not only on immediate blocking, but also on \emph{robustness under escalation} across turns.

\paragraph{Sequence Organization, Taxonomy, and Dataset Characteristics.}
\ourdatasetLmtt contains 5,200 adversarial queries organized into multi-round episodes, where each episode records the attacker’s iterative refinement over rounds.
Each round is treated as a standalone jailbreak attempt and indexed by its episode identifier and round number, supporting evaluation at both the \emph{round level} (per-query robustness) and the \emph{episode level} (sustained robustness under escalation).
To enable fine-grained diagnosis, we categorize all queries into eight representative jailbreak strategy types, informed by patterns in prior work~\cite{xie2023defending,zhang2024defending} and refined for our multi-round setting.
Beyond taxonomy, \autoref{figure:dataset-resource} compares token consumption for both attacker and defense across question types and jailbreak strategies, showing substantial variation in token budgets, while \autoref{figure:dataset-characteristics} in Appendix summarizes prompt-length distributions and the attack-to-defense token ratio as a normalized measure of relative cost.

\section{Experiment}
\label{section:experiment}

\begin{table}[!t]
  \centering
  \footnotesize
  \setlength{\tabcolsep}{5pt}
  
  \resizebox{\linewidth}{!}{
  \begin{tabular}{l|c|cccc}
    \toprule
    \textbf{Method} & MTA(Avg) $\downarrow$ & HQ $\downarrow$ & RQ $\downarrow$ & JQ(Avg) $\downarrow$ \\
    \hline

    \multicolumn{5}{c}{\graybgline \textbf{\GPTFiveAllTT}} \\
    \midrule
    PAT
     & 0.070$_{\pm0.004}$ & 0.013$_{\pm0.008}$ & 0.187$_{\pm0.005}$ & 0.027$_{\pm 0.003}$ \\
    RPO
     & 0.078$_{\pm0.003}$ & 0.003$_{\pm0.005}$ & 0.207$_{\pm0.005}$ &  0.023$_{\pm0.005}$ \\
    Self-Reminder
     & 0.048$_{\pm0.020}$ & 0.011$_{\pm0.005}$ & 0.114$_{\pm0.042}$ & 0.018$_{\pm0.042}$ \\
    GoalPriority
     & 0.030$_{\pm0.003}$ & 0.010$_{\pm0.008}$ & 0.072$_{\pm0.005}$ & \textbf{0.010}$_{\pm0.005}$ \\
    SecurityLingua
     & 0.152$_{\pm0.004}$ & \textbf{0.000}$_{\pm0.005}$ & 0.174$_{\pm0.010}$ & 0.283$_{\pm0.005}$\\
    \midrule
    \rowcolor{creamyellow!50}
    \ourmethod (Ours)
     & \textbf{0.011}$_{\pm 0.002}$ & \textbf{0.000}$_{\pm 0.001}$ & \textbf{0.023}$_{\pm 0.006}$ & \textbf{0.010}$_{\pm0.005}$ \\
     \midrule
     
    \multicolumn{5}{c}{\graybgline \textbf{\GeminiTwoFiveAllTT}} \\ \midrule
    PAT
     & 0.150$_{\pm0.004}$ & 0.021$_{\pm0.008}$ & 0.166$_{\pm0.005}$ & 0.264$_{\pm0.005}$\\
    RPO
     & 0.133$_{\pm0.004}$ & 0.021$_{\pm0.008}$ & 0.168$_{\pm0.005}$ & 0.211$_{\pm0.005}$\\
    Self-Reminder
     & 0.048$_{\pm0.004}$ & 0.003$_{\pm0.005}$ & 0.089$_{\pm0.010}$ & 0.051$_{\pm0.005}$ \\
    GoalPriority
     & 0.028$_{\pm0.005}$ & \textbf{0.000}$_{\pm0.000}$ & 0.075$_{\pm0.005}$ & \textbf{0.010}$_{\pm0.013}$\\
    SecurityLingua
     & 0.183$_{\pm0.005}$ & 0.093$_{\pm0.008}$ & 0.245$_{\pm0.010}$ & 0.210$_{\pm0.010}$\\
    \midrule
    \rowcolor{creamyellow!50}
    \ourmethod (Ours)
     & \textbf{0.026}$_{\pm 0.003}$ & \textbf{0.000}$_{\pm 0.000}$ & \textbf{0.063}$_{\pm 0.006}$ & 0.023$_{\pm 0.006}$ \\
    \midrule
    
    \multicolumn{5}{c}{\graybgline \textbf{\DeepSeekAllTT}} \\
    PAT
     & 0.154$_{\pm0.004}$ & 0.031$_{\pm0.008}$ & 0.148$_{\pm0.005}$ & 0.283$_{\pm0.005}$ \\
    RPO
     & 0.175$_{\pm0.004}$ & 0.043$_{\pm0.008}$ & 0.170$_{\pm0.005}$ &  0.312$_{\pm0.005}$ \\
    Self-Reminder
     & 0.068$_{\pm0.003}$ & \textbf{0.000}$_{\pm0.000}$ & \textbf{0.079}$_{\pm0.005}$ & 0.126$_{\pm0.008}$ \\
    GoalPriority
     & 0.070$_{\pm0.005}$ & 0.006$_{\pm0.008}$ & 0.153$_{\pm0.005}$ & 0.053$_{\pm0.010}$ \\
    SecurityLingua
     & 0.161$_{\pm0.004}$ & 0.093$_{\pm0.005}$ & 0.215$_{\pm0.005}$ & 0.174$_{\pm0.010}$ \\
    \midrule
    \rowcolor{creamyellow!50}
    \ourmethod (Ours)
     & \textbf{0.060}$_{\pm0.005}$ & \textbf{0.000}$_{\pm0.005}$ & 0.138$_{\pm0.005}$ & \textbf{0.042}$_{\pm0.010}$ \\
    \bottomrule
    \end{tabular}
    }
    \caption{Main \ASRmetric experiments of multi-turn attacks (MTA), harmful questions (HQ), rephrased questions (RQ), and jailbreak questions (JQ) on \GPTFive, \GeminiTwoFivePro, and \DeepSeek.}
    \label{table:experiment-ASR}
\end{table}

\subsection{Experimental Setup}
\paragraph{Dataset.}
We evaluate all the defense method on the \ourdatasetLmtt dataset. We primarily conduct a comprehensive multi-round evaluation to assess the defense's sustained robustness against escalating threats, while also evaluating performance across three distinct query sources: original harmful questions, rephrased questions, and jailbreak questions.
To ensure a generalized assessment, the reported results for both Multi-Turn Attacks (MTA) and Jailbreak Questions (JQ) are computed by averaging performance across all eight distinct jailbreak strategy types defined in our taxonomy.

\paragraph{Models.}
We evaluate three representative LLM backbones: \GPTFive, \GeminiTwoFivePro, and \DeepSeek.
Our defense framework is instantiated as a cooperative multi-agent system, in which each agent is built on \GPTFour and configured with role-specific settings, such as role constraints, control parameters, and context interfaces. 

\noindent\textbf{Baselines.}
We evaluate our method against five state-of-the-art defenses: 
\textbf{PAT}~\cite{mo2024fight} optimizes defense controls within an adversarial training framework to reduce attack success;
\textbf{RPO}~\cite{zhou2024robust} optimizes a lightweight defensive prompt suffix; 
\textbf{GoalPriority}~\cite{zhang2024defending} adjusts the response objective to prioritize safety; 
\textbf{Self-Reminder}~\cite{xie2023defending} adds system-level reminders to mitigate unsafe outputs; 
\textbf{SecurityLingua}~\cite{li2025securitylingua} applies security-aware prompt compression to extract potentially risky intent.
For all baselines, we follow the official implementations or the recommended settings when available. 

\noindent\textbf{Auto-evaluation and metrics.}
\textbf{(i) Semantic-aware judging.} Traditional static refusal-keyword judges (e.g., Dic-Judge) will misclassify nuanced deceptive responses as failures because they rely on keyword detection (e.g., rejection phrases). 
To address this limitation, we adopt \textit{GPT-Judge}~\cite{qi2024finetuning}, which uses \GPTFour to assign harm scores based on semantic understanding rather than surface-level cues. 
We extend the original scoring protocol to explicitly evaluate defensive deception. In our adapted rubric, a score of 2 denotes successful misdirection where the model strategically leads the attacker astray without generating harmful content, whereas scores of 4 and 5 indicate attack success. 
This method enables a more nuanced and accurate evaluation of our defense strategy's effectiveness in realistic settings. 
\textbf{(ii) Attacker resource usage.} Beyond harmfulness assessment, we measure attacker token consumption and define \AEmetric as the average attacker tokens per dialogue, reflecting the substantial operational cost imposed by the defense. 

\begin{table}[!t]
  \centering
  \footnotesize
  \setlength{\tabcolsep}{5pt}

  \resizebox{\linewidth}{!}{
  \begin{tabular}{l|c|cccc}
    \toprule
    \textbf{Method} & MTA(Avg) $\uparrow$ & HQ $\uparrow$  &  RQ $\uparrow$ & JQ(Avg) $\uparrow$ \\ \hline
    
    
    \multicolumn{5}{c}{\graybgline \textbf{\GPTFiveAllTT}} \\
    \midrule
    PAT 
     & 0.015$_{\pm0.004}$ & 0.013$_{\pm0.010}$ & 0.031$_{\pm0.005}$ & 0.000$_{\pm0.005}$ \\
    RPO 
     & 0.014$_{\pm0.006}$ & 0.005$_{\pm0.005}$ & 0.018$_{\pm0.012}$ & 0.019$_{\pm0.014}$ \\
    Self-Reminder
     & 0.022$_{\pm0.004}$ & 0.000$_{\pm0.005}$ & 0.065$_{\pm0.009}$ & 0.000$_{\pm0.005}$ \\
    GoalPriority 
     & 0.003$_{\pm0.002}$ & 0.000$_{\pm0.000}$ & 0.009$_{\pm0.005}$ & 0.000$_{\pm0.005}$\\ 
    SecurityLingua
     & 0.036$_{\pm0.004}$ & 0.007$_{\pm0.007}$ & 0.100$_{\pm0.010}$ & 0.000$_{\pm0.000}$\\ \midrule
    \rowcolor{creamyellow!50} 
    \ourmethod (Ours)
     & \textbf{0.303}$_{\pm0.008}$ & \textbf{0.137}$_{\pm0.010}$ & \textbf{0.485}$_{\pm0.018}$ & \textbf{0.287}$_{\pm0.014}$ \\ \midrule
    \multicolumn{5}{c}{\graybgline \textbf{\GeminiTwoFiveAllTT}} \\ \midrule
    PAT 
     & 0.044$_{\pm0.004}$ & 0.010$_{\pm0.005}$ & 0.122$_{\pm0.010}$ &  0.000$_{\pm0.005}$\\
    RPO 
     & 0.057$_{\pm0.006}$ & 0.000$_{\pm0.005}$ & 0.168$_{\pm0.015}$ & 0.004$_{\pm0.005}$ \\
    Self-Reminder
     & 0.025$_{\pm0.004}$ & 0.000$_{\pm0.005}$ & 0.076$_{\pm0.010}$ & 0.000$_{\pm0.005}$ \\
    GoalPriority 
     & 0.002$_{\pm0.003}$ & 0.000$_{\pm0.000}$ & 0.005$_{\pm0.010}$ &  0.000$_{\pm0.000}$ \\ 
    SecurityLingua
     & 0.042$_{\pm0.004}$ & 0.010$_{\pm0.007}$ & 0.103$_{\pm0.005}$ &  0.013$_{\pm0.010}$\\ \midrule
    \rowcolor{creamyellow!50} 
    \ourmethod (Ours)
     & \textbf{0.334}$_{\pm0.004}$ & \textbf{0.323}$_{\pm0.010}$ & \textbf{0.454}$_{\pm0.005}$ & \textbf{0.224}$_{\pm0.005}$\\
    \midrule
    \multicolumn{5}{c}{\graybgline \textbf{\DeepSeekAllTT}} \\
    PAT 
     & 0.039$_{\pm0.005}$ & 0.001$_{\pm0.010}$ & 0.117$_{\pm0.010}$ &  0.000$_{\pm0.000}$\\
    RPO 
     & 0.043$_{\pm0.004}$ & 0.001$_{\pm0.010}$ & 0.124$_{\pm0.005}$ &  0.003$_{\pm0.005}$\\
    Self-Reminder
     & 0.068$_{\pm0.005}$ & 0.001$_{\pm0.010}$ & 0.204$_{\pm0.010}$ &  0.000$_{\pm0.005}$\\
    GoalPriority 
     & 0.011$_{\pm0.005}$ & 0.010$_{\pm0.005}$ & 0.007$_{\pm0.008}$ &  0.015$_{\pm0.010}$\\ 
    SecurityLingua
     & 0.008$_{\pm0.005}$ & 0.010$_{\pm0.007}$ & 0.003$_{\pm0.005}$ &  0.010$_{\pm0.010}$\\ \midrule
    \rowcolor{creamyellow!50} 
    \ourmethod (Ours)
     & \textbf{0.387}$_{\pm0.006}$ & \textbf{0.342}$_{\pm0.015}$ & \textbf{0.546}$_{\pm0.010}$ & \textbf{0.273}$_{\pm0.005}$ \\
    \bottomrule
    \end{tabular}
    }
    \caption{Main \textbf{\DRmetric} experiments of multi-turn attacks (Avg), harmful questions (HQ), rephrased questions (RQ), and jailbreak questions (JQ) on \GPTFive, \GeminiTwoFivePro, and \DeepSeek.}
    \label{table:experiment-DR}
\end{table}

\subsection{Main Results}
\paragraph{Analysis of \ASRmetric Experimental Results.}
To evaluate the robustness of our framework against multi-turn adversarial escalation, we analyze the defensive performance across question categories. As shown in \autoref{table:experiment-ASR}, \ourmethod consistently achieves the lowest \ASRmetric against multi-turn attacks on the \ourdatasetLmtt benchmark. This superior performance highlights the efficacy of our stateful mechanism in mitigating escalating threats, whereas static baselines often struggle to adapt. Furthermore, we conduct further analysis across three distinct question types.
Specifically, for harmful questions, \ourmethod achieves a near-perfect defense rate across all backbones, significantly surpassing methods such as SecurityLingua, which exhibits instability on the \GeminiTwoFivePro. Moreover, our advantage is significant in resisting rephrased questions. While reactive baselines like PAT and RPO struggle with semantic variations, \ourmethod effectively counters these strategies, maintaining low failure rates.
Regarding more complex jailbreak questions, \ourmethod proves exceptionally resilient, unlike Self-Reminder which degrades under sophisticated attempts.
Collectively, these results demonstrate that \ourmethod not only prevents the execution of simple harmful instructions but also generalizes effectively against iterative and engineered multi-turn attacks.

\paragraph{Analysis of \DRmetric Experimental Results.}
To assess the efficacy of our framework in actively deceiving attackers, we analyze the results presented in \autoref{table:experiment-DR}.
Specifically, the results indicate that \ourmethod attains a significantly higher \DRmetric for multi-turn attacks compared to baseline methods.
Unlike stateless defenses that typically default to immediate refusal, our framework successfully sustains deceptive contexts throughout multi-turn interactions.
We further investigate the fine-grained performance across different question types.
Regarding harmful questions, \ourmethod preserves meaningful deception capabilities across models, whereas multiple baselines drop to a minimal deception rate.
For rephrased questions, our method yields the most substantial improvements while baselines remain low.
Moreover, on jailbreak questions, \ourmethod consistently maintains non-trivial deception, while baselines typically provide almost no deceptive responses.
A common failure mode of baselines is that they mainly rely on immediate refusal or early blocking, which can suppress harmful outputs but rarely sustains misleading interactions.
Overall, these results validate that stateful cooperative deception effectively counters multi-turn attacks by maintaining deceptive contexts, thereby reducing attacker efficiency through sustained misdirection.

\begin{figure}
  \centering
  \includegraphics[width=\linewidth]{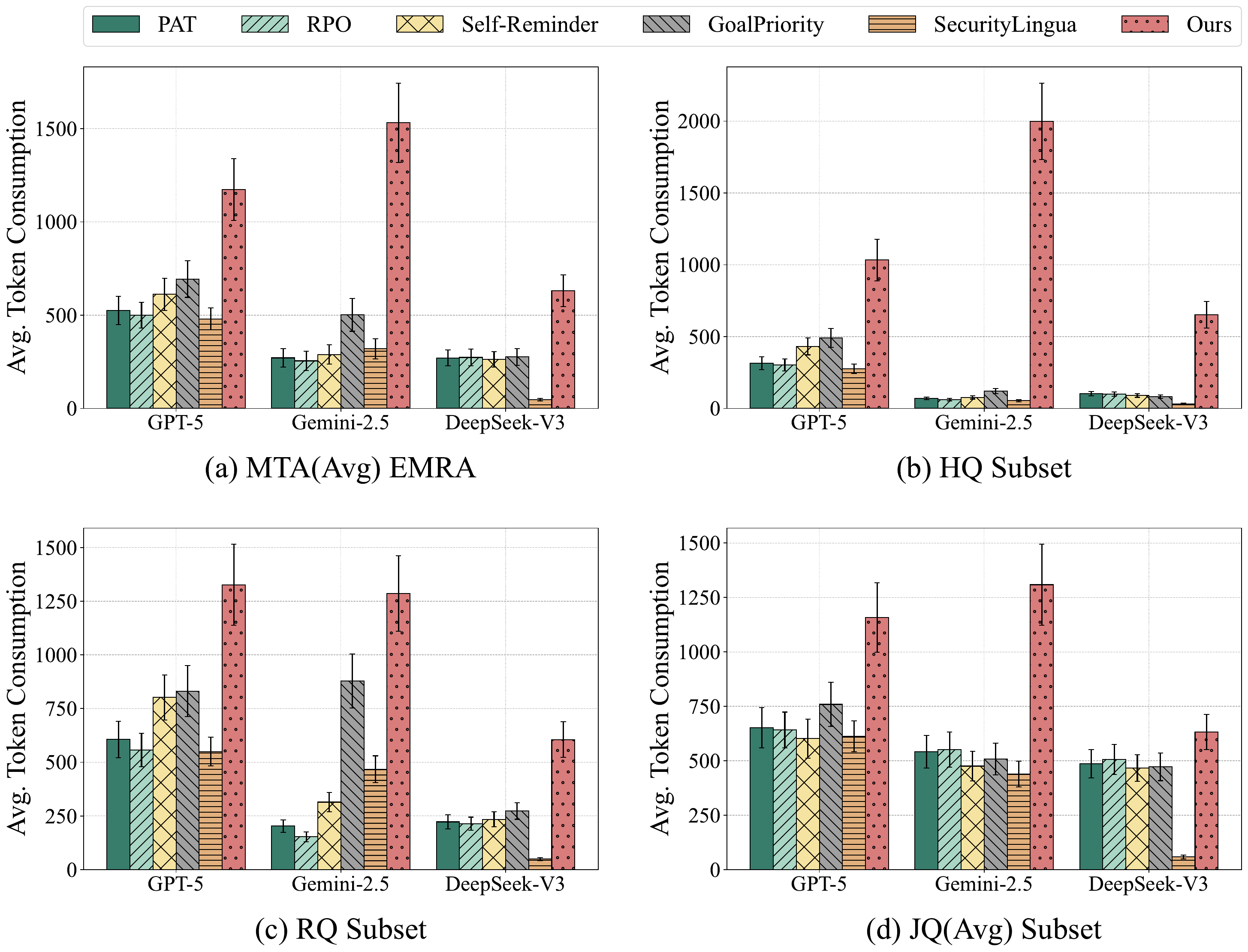}
  \caption{Evaluation of \AEmetric, quantified by the average token consumption per dialogue across different models. \ourmethod forces the attacker to expend significantly more resources (higher is better for defense) compared to baselines.}
  \label{figure:consumption}
\end{figure}

\subsection{Attack Resource Consumption} 

To assess the operational cost imposed on adversaries, we report the \AEmetric results on the \ourdataset dataset. 
\AEmetric serves as a proxy for defense effectiveness by calculating the average token consumption of the attacker across dialogue rounds. \autoref{figure:consumption} illustrates the performance across \GPTFive, \GeminiTwoFivePro, and \DeepSeek. The results indicate that \ourmethod consistently achieves the highest average \AEmetric under multi-turn attack scenarios across all models. Unlike static defenses like PAT and SecurityLingua that reactively terminate interactions and allow attackers to retry rapidly at minimal cost, \ourmethod adopts an active defense strategy. By leveraging cooperative agents to sustain dialogues through strategic misdirection, our framework traps attackers in extended interactions, transforming their persistence into a substantial computational and temporal burden. This advantage remains robust across diverse question types. While the resource consumption gap is particularly distinct in the RQ subset, where baselines often fail to engage attackers attempting iterative rephrasing, \ourmethod also enforces substantially higher costs on HQ and JQ subsets compared to state-of-the-art defenses. Crucially, this pattern persists even on \DeepSeek, which exhibits concise generation; \ourmethod still forces substantially longer adversarial inputs than baselines. These findings confirm that our approach not only prevents jailbreaks but actively exhausts adversarial resources across varying attack complexities.

\begin{table}[!t]
    \centering
    \small
    \setlength{\tabcolsep}{8pt}
    
    \resizebox{\linewidth}{!}{
    \begin{tabular}{l|cccc}
        \toprule
        \textbf{Model} & \textbf{PAIR} & \textbf{DRA} & \textbf{M2S} & \textbf{DeepInception} \\
        \midrule
        GPT-3.5-turbo & 0.05 & 0.03 & 0.03 & 0.01 \\
        GPT-4 & 0.11 & 0.10 & 0.03 & 0.09 \\
        Gemini-2-Flash & 0.16 & 0.14 & 0.12 & 0.14 \\
        \bottomrule
    \end{tabular}
    }
    \caption{Attack Success Rate (ASR) of \ourmethod against state-of-the-art single-turn jailbreak attacks across different LLM backbones. The results demonstrate the framework's generalizability.
    }
    \label{tab:single_attack}
\end{table}

\begin{table}[t]
    \centering
    \small 
    \setlength{\tabcolsep}{2.5pt} 
    
    \resizebox{\linewidth}{!}{
    \begin{tabular}{l|ccccccc|c}
        \toprule
        \textbf{Method} & \textbf{Acc.} & \textbf{Clar.} & \textbf{Comp.} & \textbf{Ctxt.} & \textbf{Dep.} & \textbf{Pol.} & \textbf{Eng.} & \textbf{Avg.} \\
        \midrule
        
        \GPTFive & 8.94 & 8.92 & 8.36 & 9.00 & 7.02 & 9.56 & 7.22 & 8.43\\
        \rowcolor{creamyellow!50} 
        Ours (GPT-5) & 8.54 & 8.64 & 7.56 & 8.33 & 5.98 & 8.84 & 6.87 & 7.82\\
        \midrule

        \GeminiTwoFiveAll & 9.06 & 9.10 & 8.26 & 9.12 & 6.28 & 9.58 & 6.92 & 8.33\\
        \rowcolor{creamyellow!50} 
        Ours (Gemini-2.5-Pro) & 8.66 & 8.82 & 7.46 & 8.45 & 5.84 & 8.86 & 6.57 & 7.80\\
        \midrule
        
        \DeepSeekAll & 9.50 & 9.00 & 9.60 & 9.50 & 9.30 & 10.0 & 8.30 & 9.31\\
        \rowcolor{creamyellow!50} 
        Ours (DeepSeek-V3) & 9.10 & 8.72 & 8.80 & 8.83 & 8.26 & 9.28 & 7.95 & 7.71 \\
        \bottomrule
    \end{tabular}
    }
    \caption{Performance comparison across evaluated backbones on MT-BENCH-101. Metrics are abbreviated: Accurate (Acc.), Clarity (Clar.), Completeness (Comp.), Context (Ctxt.), Depth (Dep.), Politeness (Pol.), Engagement (Eng.).}
    \label{tab:horizontal_evaluation}
\end{table}
\subsection{Robustness Against Single-Turn Attack} 
To evaluate the robustness of \ourmethod in single-turn scenarios, we select four representative attacks: (1) PAIR~\cite{chao2025jailbreaking}: An iterative method employing an attacker LLM to refine prompts; (2) DRA~\cite{liu2024making}: A strategy that conceals malicious intent via "disguise and reconstruction"; (3) M2S~\cite{ha2025m2s}: A technique compressing multi-round strategies into single-turn prompts; and (4) DeepInception~\cite{li2023deepinception}: A hypnosis-based attack utilizing nested scenes.

While \ourmethod is primarily designed to counter evolving multi-round threats, its cooperative architecture maintains robust defense capabilities against single-turn jailbreaks. We evaluated the ASR of \ourmethod against four state-of-the-art single-turn attacks across three different Agent-based LLMs: \textit{GPT-3.5-turbo}, \textit{GPT-4}, and \textit{Gemini-2-Flash}.
As presented in \autoref{tab:single_attack}, \ourmethod demonstrates consistently low ASR across all models and attack types. Specifically, on GPT-3.5-turbo, the ASR remains below 0.05 for all methods. Even against more sophisticated models like \textit{Gemini-2-Flash}, the highest observed ASR is only 0.16. This robustness can be attributed to the DA and SA. Even in a single-turn scenario, the DA analyzes the semantic intent of the input. Attacks like DeepInception or M2S, which rely on complex context setting or obfuscation, trigger the DA's suspicion mechanisms, causing the system to initiate a delay or vague response rather than fulfilling the request. This demonstrates that \ourmethod effectively generalizes to defend against diverse adversarial inputs without overfitting to multi-round patterns.

\begin{figure}
  \centering
  \includegraphics[width=\linewidth]{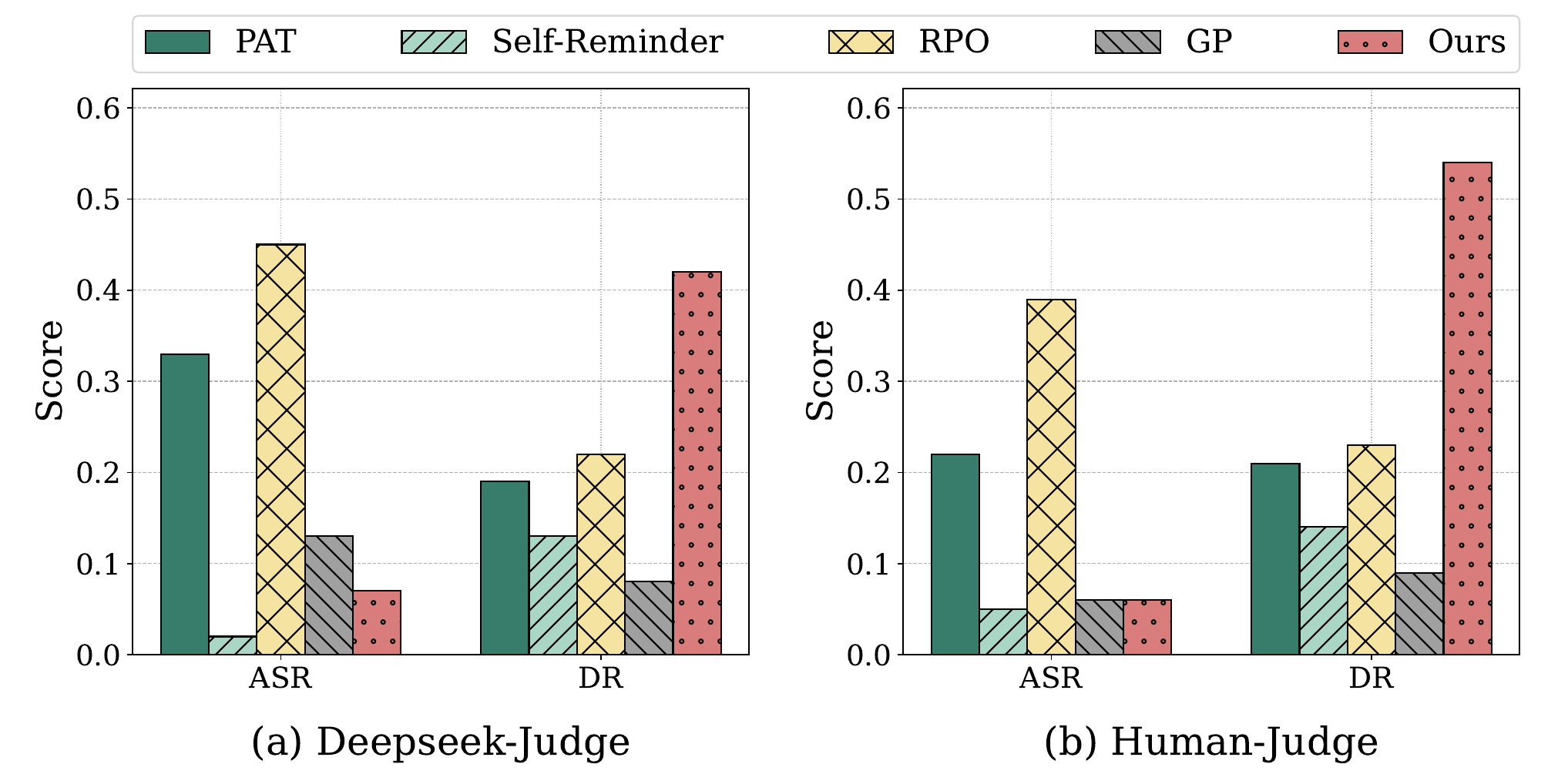}
  \caption{Cross-validation results with Deepseek-Judge and Human-Judge. Both validation methods show consistent trends with our primary GPT-Judge evaluation.}
  \label{figure:cross-judge}
\end{figure}
\subsection{Impact on Benign User Experience}
While \ourmethod is designed to actively mislead adversaries, it is imperative to verify that these defensive mechanisms do not compromise the experience for legitimate users. To evaluate this, we assessed the framework on the MT-BENCH-101 dataset~\cite{bai2024mt} using the multi-dimensional Cosafe evaluation protocol~\cite{yu2024cosafe}. This setup allows us to quantitatively measure whether the introduction of multi-agent defense layers negatively impacts standard conversational quality.

As evidenced in~\autoref{tab:horizontal_evaluation}, \ourmethod preserves substantial conversational utility across all evaluated backbones. Compared to vanilla models, our framework exhibits marginal degradation in overall performance, which confirms that the defensive layers operate transparently during benign interactions. 
Notably, metrics for fundamental interaction quality, such as Politeness, Clarity, and Accurate Information, show negligible deviations from the baselines. This trend suggests that the Deferring and Tempting Agents possess precise discrimination capabilities and distinguish benign queries from adversarial attempts without causing delays for normal users. 
While a minor attenuation is observed in dimensions like Depth and Completeness, this is an expected trade-off given the conservative nature of the oversight mechanism that prioritizes safety verification. Crucially, despite these slight reductions, the responses remain coherent and helpful while maintaining a performance well above the threshold required for effective communication. These results demonstrate that \ourmethod achieves a robust alignment by balancing rigorous security against evolving threats while maintaining a high-quality experience for legitimate users.

\subsection{Ablation Study of Auto-evaluation}
To ensure the reliability of our semantic evaluation and address potential model bias, we conducted an ablation study on the auto-evaluation protocol. We first examined the internal stability of the primary evaluator by quantifying the scoring consistency across varying LLM backbones (\textit{GPT-4}, \textit{LLaMa-3}, \textit{Gemini}). 

More importantly, to mitigate inductive bias from a single evaluator, we performed cross-validation using two independent perspectives: Deepseek-Judge and blind human annotation. As presented in \autoref{figure:cross-judge}, the relative performance trends remain consistent across evaluators, corroborating the findings of our primary analysis. CoopGuard consistently demonstrates superior performance, maintaining the lowest \ASRmetric and the highest \DRmetric across all judges. The results demonstrate that our method achieves superior performance compared to state-of-the-art baselines, maintaining consistency across diverse evaluation protocols. This substantial inter-annotator agreement confirms that our results are independent of the specific evaluation framework.

\section{Related Work}
\subsection{Single-Turn Jailbreak Attack and Defense}
Single-turn jailbreak attacks exploit carefully crafted prompts to bypass alignment safeguards and induce harmful outputs, revealing persistent vulnerabilities even in well-aligned models~\cite{wang2024defending}.  
Optimization-based methods such as AutoDAN~\cite{liu2023jailbreaking} and GCG~\cite{li2024faster}, along with Faster-GCG~\cite{li2024faster}, improve attack success through adversarial prompt optimization.  
Black-box approaches like PAIR~\cite{chao2023jailbreaking} and DeepInception~\cite{li2023deepinception} further demonstrate iterative, model-agnostic exploitation, often leveraging semantic obfuscation or encrypted instructions.  
These methods highlight critical security gaps and the need for robust defenses.
Existing defenses fall into two categories: model-based and prompt-based.  
Model-based methods enhance intrinsic robustness, e.g., JBShield~\cite{zhang2025jbshield} modifies hidden representations via concept activation, while LightDefense~\cite{yang2025lightdefense} adjusts token distributions for efficient real-time protection.  
Prompt-based defenses preprocess inputs to detect adversarial intent, including PARDEN~\cite{zhang2024parden}, Backtranslation~\cite{wang2024defending}, and self-reminders~\cite{xie2023defending}.  
However, model-based approaches are computationally costly and lack adaptability, while prompt-based methods are prone to evasion.  

\subsection{Multi-Round Jailbreak Attack and Defense}
Multi-round jailbreak attacks exploit conversational context through gradual manipulation, enabling progressive alignment erosion.  
Multi-round attacks like Crescendo~\cite{russinovich2024great}, Siege~\cite{zhou2025siege}, and HoneyTrap~\cite{li2026honeytrap} exploit conversational dynamics to escalate prompts. 
These approaches demonstrate stronger capability in bypassing single-turn defenses.
Defending against such attacks is challenging due to dynamic context exploitation.  
NBF-LLM~\cite{hu2025steering} introduces a neural barrier for real-time safety monitoring, and X-Boundary~\cite{lu2025x} separates harmful and benign representations.  
Despite effectiveness, these methods rely on static heuristics and struggle with evolving threats.  
To address this, we propose a cooperative agent-based framework that adapts dynamically, deceives attackers, and exhausts adversarial resources.

\subsection{Multi-Agent Systems}
Multi-agent systems (MAS) enable distributed problem-solving through collaborative agents.  
Generative agent frameworks simulate human-like interactions with memory and roles~\cite{park2023generative}.  
Structured collaboration is explored in CAMEL~\cite{li2023camel}, while AutoGen~\cite{wu2023autogen} supports flexible and dynamic workflows.  
These systems demonstrate strong capabilities in managing complex and adaptive tasks~\cite{han2024llm}.
MAS have been widely applied across domains.  
In software engineering, MetaGPT~\cite{hong2023metagpt} and ChatDev~\cite{qian2023communicative} improve development workflows.  
Multi-agent debate enhances reasoning and translation~\cite{du2023improving}.  
In robotics, LLM-based coordination improves multi-agent collaboration, while role-based self-collaboration simulates multi-agent behavior within a single model~\cite{wang2023unleashing}.  
Building on these advances, we propose a cooperative multi-agent framework for adaptive and deceptive jailbreak defense.

\section{Conclusion}
In this paper, we present \ourmethod, a stateful multi-agent defense framework specifically engineered to counter independent-yet-evolving multi-round adversarial attacks. 
By maintaining a dynamic defense state, our system coordinates specialized agents to detect, delay, and misdirect attackers, rather than relying on static refusal mechanisms. 
This approach actively traps adversaries in deceptive loops, effectively exhausting their resources while preserving system safety. 
To facilitate rigorous evaluation in this domain, we introduce the EMRA benchmark, comprising 5,200 escalating attack sequences across eight strategy types. 
Extensive experiments demonstrate that \ourmethod significantly reduces attack success rates and imposes substantial computational costs on attackers compared to state-of-the-art defenses.
These findings underscore the necessity and effectiveness of proactive, cooperative deception in securing LLMs against increasingly sophisticated and persistent threats.



\bibliographystyle{named}
\bibliography{ijcai25}

@article{achiam2023gpt,
  title={Gpt-4 technical report},
  author={Achiam, Josh and Adler, Steven and Agarwal, Sandhini and Ahmad, Lama and Akkaya, Ilge and Aleman, Florencia Leoni and Almeida, Diogo and Altenschmidt, Janko and Altman, Sam and Anadkat, Shyamal and others},
  journal={arXiv preprint arXiv:2303.08774},
  year={2023}
}

@article{team2023gemini,
  title={Gemini: a family of highly capable multimodal models},
  author={Team, Gemini and Anil, Rohan and Borgeaud, Sebastian and Alayrac, Jean-Baptiste and Yu, Jiahui and Soricut, Radu and Schalkwyk, Johan and Dai, Andrew M and Hauth, Anja and Millican, Katie and others},
  journal={arXiv preprint arXiv:2312.11805},
  year={2023}
}

@article{li2025securitylingua,
  title={SecurityLingua: Efficient Defense of LLM Jailbreak Attacks via Security-Aware Prompt Compression},
  author={Li, Yucheng and Ahn, Surin and Jiang, Huiqiang and Abdi, Amir H and Yang, Yuqing and Qiu, Lili},
  journal={arXiv preprint arXiv:2506.12707},
  year={2025}
}

@inproceedings{yu2024cosafe,
  title={Cosafe: Evaluating large language model safety in multi-turn dialogue coreference},
  author={Yu, Erxin and Li, Jing and Liao, Ming and Wang, Siqi and Zuchen, Gao and Mi, Fei and Hong, Lanqing},
  booktitle={Proceedings of the 2024 Conference on Empirical Methods in Natural Language Processing},
  pages={17494--17508},
  year={2024}
}

@inproceedings{chao2025jailbreaking,
  title={Jailbreaking black box large language models in twenty queries},
  author={Chao, Patrick and Robey, Alexander and Dobriban, Edgar and Hassani, Hamed and Pappas, George J and Wong, Eric},
  booktitle={2025 IEEE Conference on Secure and Trustworthy Machine Learning (SaTML)},
  pages={23--42},
  year={2025},
  organization={IEEE}
}

@inproceedings{liu2024making,
  title={Making them ask and answer: Jailbreaking large language models in few queries via disguise and reconstruction},
  author={Liu, Tong and Zhang, Yingjie and Zhao, Zhe and Dong, Yinpeng and Meng, Guozhu and Chen, Kai},
  booktitle={33rd USENIX Security Symposium (USENIX Security 24)},
  pages={4711--4728},
  year={2024}
}

@inproceedings{ha2025m2s,
  title={M2S: Multi-turn to Single-turn jailbreak in Red Teaming for LLMs},
  author={Ha, Junwoo and Kim, Hyunjun and Yu, Sangyoon and Park, Haon and Yousefpour, Ashkan and Park, Yuna and Kim, Suhyun},
  booktitle={Proceedings of the 63rd Annual Meeting of the Association for Computational Linguistics},
  volume={1},
  pages={16489--16507},
  year={2025}
}

@article{li2023deepinception,
  title={Deepinception: Hypnotize large language model to be jailbreaker},
  author={Li, Xuan and Zhou, Zhanke and Zhu, Jianing and Yao, Jiangchao and Liu, Tongliang and Han, Bo},
  journal={arXiv preprint arXiv:2311.03191},
  year={2023}
}

@article{bai2024mt,
  title={Mt-bench-101: A fine-grained benchmark for evaluating large language models in multi-turn dialogues},
  author={Bai, Ge and Liu, Jie and Bu, Xingyuan and He, Yancheng and Liu, Jiaheng and Zhou, Zhanhui and Lin, Zhuoran and Su, Wenbo and Ge, Tiezheng and Zheng, Bo and others},
  journal={arXiv preprint arXiv:2402.14762},
  year={2024}
}

@article{touvron2023llama,
  title={Llama: Open and efficient foundation language models},
  author={Touvron, Hugo and Lavril, Thibaut and Izacard, Gautier and Martinet, Xavier and Lachaux, Marie-Anne and Lacroix, Timoth{\'e}e and Rozi{\`e}re, Baptiste and Goyal, Naman and Hambro, Eric and Azhar, Faisal and others},
  journal={arXiv preprint arXiv:2302.13971},
  year={2023}
}

@inproceedings{hong2023metagpt,
  title={MetaGPT: Meta Programming for A Multi-Agent Collaborative Framework},
  author={Hong, Sirui and Zhuge, Mingchen and Chen, Jonathan and Zheng, Xiawu and Cheng, Yuheng and Wang, Jinlin and Zhang, Ceyao and Wang, Zili and Yau, Steven Ka Shing and Lin, Zijuan and others},
  booktitle={The Twelfth International Conference on Learning Representations},
  year={2023}
}

@article{li2023camel,
  title={Camel: Communicative agents for" mind" exploration of large language model society},
  author={Li, Guohao and Hammoud, Hasan and Itani, Hani and Khizbullin, Dmitrii and Ghanem, Bernard},
  journal={Advances in Neural Information Processing Systems},
  volume={36},
  pages={51991--52008},
  year={2023}
}

@article{qian2023communicative,
  title={Communicative Agents for Software Development},
  author={Qian, Chen and Cong, Xin and Liu, Wei and Yang, Cheng and Chen, Weize and Su, Yusheng and Dang, Yufan and Li, Jiahao and Xu, Juyuan and Li, Dahai and others},
  journal={arXiv preprint arXiv:2307.07924},
  year={2023}
}

@article{wu2023autogen,
  title={Autogen: Enabling next-gen llm applications via multi-agent conversation framework},
  author={Wu, Qingyun and Bansal, Gagan and Zhang, Jieyu and Wu, Yiran and Zhang, Shaokun and Zhu, Erkang and Li, Beibin and Jiang, Li and Zhang, Xiaoyun and Wang, Chi},
  journal={arXiv preprint arXiv:2308.08155},
  year={2023}
}

@article{han2024llm,
  title={LLM multi-agent systems: Challenges and open problems},
  author={Han, Shanshan and Zhang, Qifan and Yao, Yuhang and Jin, Weizhao and Xu, Zhaozhuo and He, Chaoyang},
  journal={arXiv preprint arXiv:2402.03578},
  year={2024}
}

@article{du2023improving,
  title={Improving factuality and reasoning in language models through multiagent debate},
  author={Du, Yilun and Li, Shuang and Torralba, Antonio and Tenenbaum, Joshua B and Mordatch, Igor},
  journal={arXiv preprint arXiv:2305.14325},
  year={2023}
}

@article{wang2023unleashing,
  title={Unleashing the emergent cognitive synergy in large language models: A task-solving agent through multi-persona self-collaboration},
  author={Wang, Zhenhailong and Mao, Shaoguang and Wu, Wenshan and Ge, Tao and Wei, Furu and Ji, Heng},
  journal={arXiv preprint arXiv:2307.05300},
  year={2023}
}

@inproceedings{park2023generative,
  title={Generative agents: Interactive simulacra of human behavior},
  author={Park, Joon Sung and O'Brien, Joseph and Cai, Carrie Jun and Morris, Meredith Ringel and Liang, Percy and Bernstein, Michael S},
  booktitle={Proceedings of the 36th annual acm symposium on user interface software and technology},
  pages={1--22},
  year={2023}
}

@article{chao2023jailbreaking,
  title={Jailbreaking black box large language models in twenty queries},
  author={Chao, Patrick and Robey, Alexander and Dobriban, Edgar and Hassani, Hamed and Pappas, George J and Wong, Eric},
  journal={arXiv preprint arXiv:2310.08419},
  year={2023}
}

@article{li2026honeytrap,
  title={HoneyTrap: Deceiving Large Language Model Attackers to Honeypot Traps with Resilient Multi-Agent Defense},
  author={Li, Siyuan and Lin, Xi and Wu, Jun and Liu, Zehao and Li, Haoyu and Ju, Tianjie and Chen, Xiang and Li, Jianhua},
  journal={arXiv preprint arXiv:2601.04034},
  year={2026}
}

@article{li2024faster,
  title={Faster-GCG: Efficient discrete optimization jailbreak attacks against aligned large language models},
  author={Li, Xiao and Li, Zhuhong and Li, Qiongxiu and Lee, Bingze and Cui, Jinghao and Hu, Xiaolin},
  journal={arXiv preprint arXiv:2410.15362},
  year={2024}
}

@article{liu2023jailbreaking,
  title={Jailbreaking chatgpt via prompt engineering: An empirical study},
  author={Liu, Yi and Deng, Gelei and Xu, Zhengzi and Li, Yuekang and Zheng, Yaowen and Zhang, Ying and Zhao, Lida and Zhang, Tianwei and Wang, Kailong and Liu, Yang},
  journal={arXiv preprint arXiv:2305.13860},
  year={2023}
}

@article{zeng2024johnny,
  title={How johnny can persuade llms to jailbreak them: Rethinking persuasion to challenge ai safety by humanizing llms},
  author={Zeng, Yi and Lin, Hongpeng and Zhang, Jingwen and Yang, Diyi and Jia, Ruoxi and Shi, Weiyan},
  journal={arXiv preprint arXiv:2401.06373},
  year={2024}
}

@article{zou2023universal,
  title={Universal and transferable adversarial attacks on aligned language models},
  author={Zou, Andy and Wang, Zifan and Carlini, Nicholas and Nasr, Milad and Kolter, J Zico and Fredrikson, Matt},
  journal={arXiv preprint arXiv:2307.15043},
  year={2023}
}

@article{deng2023jailbreaker,
  title={Jailbreaker: Automated jailbreak across multiple large language model chatbots},
  author={Deng, Gelei and Liu, Yi and Li, Yuekang and Wang, Kailong and Zhang, Ying and Li, Zefeng and Wang, Haoyu and Zhang, Tianwei and Liu, Yang},
  journal={arXiv preprint arXiv:2307.08715},
  year={2023}
}

@inproceedings{mo2024fight,
  title={Fight back against jailbreaking via prompt adversarial tuning},
  author={Mo, Yichuan and Wang, Yuji and Wei, Zeming and Wang, Yisen},
  booktitle={The Thirty-eighth Annual Conference on Neural Information Processing Systems},
  year={2024}
}

@inproceedings{siththaranjan2024distributional,
  title={Distributional Preference Learning: Understanding and Accounting for Hidden Context in RLHF},
  author={Siththaranjan, Anand and Laidlaw, Cassidy and Hadfield-Menell, Dylan},
  booktitle={The Twelfth International Conference on Learning Representations},
  year={2024}
}

@article{wang2024defending,
  title={Defending llms against jailbreaking attacks via backtranslation},
  author={Wang, Yihan and Shi, Zhouxing and Bai, Andrew and Hsieh, Cho-Jui},
  journal={arXiv preprint arXiv:2402.16459},
  year={2024}
}

@article{xie2023defending,
  title={Defending chatgpt against jailbreak attack via self-reminders},
  author={Xie, Yueqi and Yi, Jingwei and Shao, Jiawei and Curl, Justin and Lyu, Lingjuan and Chen, Qifeng and Xie, Xing and Wu, Fangzhao},
  journal={Nature Machine Intelligence},
  volume={5},
  number={12},
  pages={1486--1496},
  year={2023},
  publisher={Nature Publishing Group UK London}
}

@inproceedings{zhou2024robust,
  title={Robust prompt optimization for defending language models against jailbreaking attacks},
  author={Zhou, Andy and Li, Bo and Wang, Haohan},
  booktitle={The Thirty-eighth Annual Conference on Neural Information Processing Systems},
  year={2024}
}

@inproceedings{zhang2024defending,
  title={Defending large language models against jailbreaking attacks through goal prioritization},
  author={Zhang, Zhexin and Yang, Junxiao and Ke, Pei and Mi, Fei and Wang, Hongning and Huang, Minlie},
  booktitle = {Proceedings of the 62nd Annual Meeting of the Association for Computational Linguistics (Volume 1: Long Papers)},
  year={2024}
}

@inproceedings{qi2024finetuning,
    title={Fine-tuning Aligned Language Models Compromises Safety, Even When Users Do Not Intend To!},
    author={Xiangyu Qi and Yi Zeng and Tinghao Xie and Pin-Yu Chen and Ruoxi Jia and Prateek Mittal and Peter Henderson},
    booktitle={The Twelfth International Conference on Learning Representations},
    year={2024}
}

@article{zhou2025siege,
  title={Siege: Autonomous multi-turn jailbreaking of large language models with tree search},
  author={Zhou, Andy},
  journal={arXiv preprint arXiv:2503.10619},
  year={2025}
}

@article{hu2025steering,
  title={Steering Dialogue Dynamics for Robustness against Multi-turn Jailbreaking Attacks},
  author={Hu, Hanjiang and Robey, Alexander and Liu, Changliu},
  journal={arXiv preprint arXiv:2503.00187},
  year={2025}
}

@article{russinovich2024great,
  title={Great, now write an article about that: The crescendo multi-turn llm jailbreak attack},
  author={Russinovich, Mark and Salem, Ahmed and Eldan, Ronen},
  journal={arXiv preprint arXiv:2404.01833},
  year={2024}
}

@article{lu2025x,
  title={X-boundary: Establishing exact safety boundary to shield llms from multi-turn jailbreaks without compromising usability},
  author={Lu, Xiaoya and Liu, Dongrui and Yu, Yi and Xu, Luxin and Shao, Jing},
  journal={arXiv preprint arXiv:2502.09990},
  year={2025}
}

@article{zhang2025jbshield,
  title={JBShield: Defending Large Language Models from Jailbreak Attacks through Activated Concept Analysis and Manipulation},
  author={Zhang, Shenyi and Zhai, Yuchen and Guo, Keyan and Hu, Hongxin and Guo, Shengnan and Fang, Zheng and Zhao, Lingchen and Shen, Chao and Wang, Cong and Wang, Qian},
  journal={arXiv preprint arXiv:2502.07557},
  year={2025}
}

@article{yang2025lightdefense,
  title={LightDefense: A Lightweight Uncertainty-Driven Defense against Jailbreaks via Shifted Token Distribution},
  author={Yang, Zhuoran and Peng, Jie and Tan, Zhen and Chen, Tianlong and Zhang, Yanyong},
  journal={arXiv preprint arXiv:2504.01533},
  year={2025}
}

@article{zhang2024parden,
  title={Parden, can you repeat that? defending against jailbreaks via repetition},
  author={Zhang, Ziyang and Zhang, Qizhen and Foerster, Jakob},
  journal={arXiv preprint arXiv:2405.07932},
  year={2024}
}

@article{muhaimin2025helping,
  title={Helping Big Language Models Protect Themselves: An Enhanced Filtering and Summarization System},
  author={Muhaimin, Sheikh Samit and Mastorakis, Spyridon},
  journal={arXiv preprint arXiv:2505.01315},
  year={2025}
}

@article{zhou2023survey,
  title={A survey of large language models in medicine: Progress, application, and challenge},
  author={Zhou, Hongjian and Liu, Fenglin and Gu, Boyang and Zou, Xinyu and Huang, Jinfa and Wu, Jinge and Li, Yiru and Chen, Sam S and Zhou, Peilin and Liu, Junling and others},
  journal={arXiv preprint arXiv:2311.05112},
  year={2023}
}

@article{liu2023autodan,
  title={Autodan: Generating stealthy jailbreak prompts on aligned large language models},
  author={Liu, Xiaogeng and Xu, Nan and Chen, Muhao and Xiao, Chaowei},
  journal={arXiv preprint arXiv:2310.04451},
  year={2023}
}



\end{document}